# Combining Particle Tracking with Electromagnetic Radiation Showers: Merging GPT and Geant4 with Visualization

David H. Dowell[1], Munther Hindi[2], S.B. van der Geer[3] and M. J. de Loos[3]
[1]SLAC Accelerator Center
[2]Independent Researcher
[3]Pulsar Physics

**Abstract**

Field emitted electrons can seriously affect the operation of high-field, high-duty factor electron accelerators. Accelerated field emission can result in high average power beams which can radiation damage beamline components. In addition, localized losses generate thermal hot spots whose outgassing degrades the ultra-high vacuum required in photoinjectors and cryomodules. However, despite their importance, the effects of field emission are rarely included in the design and engineering of electron injectors. This work attempts to remedy this situation by combining two well-known and well-documented programs, GPT and Geant4, to track electrons and their losses in an injector beamline. This paper describes a system of programs which simulates electron paths and losses along the beamline. In addition, the tracking results can be zoomed and panned along and about the beampipe envelope using an open-source 3D CAD program. The scattering albedo calculations of the combined program, GPT-Geant4, are shown to be in good agreement with the literature albedos. The paper concludes with a dark current simulation for the LCLS-II injector from the cathode to the collimator at 1.5 m from the cathode.

### I. Introduction.

As state-of-the-art photocathode electron injectors evolve toward continuous wave (CW) operation, understanding the sources and transport of beam halo and dark current becomes increasingly important. CW operation at high accelerating fields leads to dark current beams with high average power which can activate beamline components and damage cryogenic accelerators. Therefore, we need to better quantify the dark current and to specify its sources and its losses in the beamline. This publication presents a new simulation workflow to address this issue.

Accurately simulating dark current requires modeling the source geometry and emission with micron-scale resolution and tracking the emitted electrons through the applied rf and magnetic fields on the meter-scale. Unlike a photoelectron source, field emitters are typically off axis and emit long electron bunches with large energy spread, thus the dark current is overfocused and mis-steered by the electron optics and much of it is lost somewhere along the beamline. As a radiation source, it is therefore important to quantify the location and power density of this loss to inform the design and to establish the safe operating envelope for CW, high-field injectors. Scattering of the primary dark current leads to secondary electrons which also need to be tracked because they too can be problematic.

The approach described here combines two well-established codes, GPT and Geant4, to track and scatter electrons through the beamline optics and from any obstructions. For the tracking code we chose GPT [1] because GPT already implements surface boundaries and scatters electrons using an internal scattering model. GPT is attractive to use since it provides user-defined scattering models. Implementing a custom scattering model was essential since GPT's internal scattering model is very rudimentary. Geant4 [2] was picked for the scattering code since it is widely used and is compatible with GPT's C-programming.

The combined application of GPT and Geant4, referred to GPT+Geant4, contains all the essential beam-related physics and radiation effects of an operating accelerator. It gives the accelerator physicist detailed information about how a given beamline design generates radiation and how modifying the design could mitigate it.

In addition, by connecting the GPT+Geant4 inputs to the accelerator control system's process variables (PVs), one would have a simulator which mirrors the accelerator in near real time. The operators would have accurate simulations for not only the beam's optics but also the radiation and beamline heating it generates. Such a simulator would be able to include the geometry of the inner walls of the beampipe and give the distribution of heating and local temperature rise due to beam losses under different operating conditions. From this temperature rise, the simulator would compute the outgassing rate of the beamline being heated and estimate the rise in vacuum pressure or keep a record of the radiation dose in the cryomodules.

A major goal was to combine electron optics, scattering and radiation in a unified accelerator code which is user-friendly and accurate. This was done by integrating electron scattering processes into a commonly used tracking code, thereby making radiation calculations an integral part of the accelerator design and optimization process without needing a radiation physicist or having to learn the details an EM shower code like EGS4 or Geant4.

This paper is organized into seven sections. Section II describes how the GPT and Geant4 codes were combined and function together, and how GPT+Geant4 simulations produce CAD images of the primary and secondary electron trajectories within the beampipe. The basic features of secondary emission and the GPT+Geant4 scattering model energy range are discussed in Section III. Section IV defines the electron scattering albedo and describes computing it using GPT+Geant4. In Section V the GPT+Geant4 calculation of the albedo is validated against the published experimental and simulated albedos. Comparisons of the albedo are made vs. incident energy, angle-of-incidence, and atomic number. Section VI describes the visualization and analysis tools available in the GPT+Geant4 system as illustrated by simulations for the LCLS-II injector. Section VII ends the paper with the summary and conclusions.

## II. Description of the GPT+Geant4 System.

The GPT+Geant4 simulation system consists of three parts. The first reads the GPT input file containing the surface boundary description and converts the GPT surfaces into 3D volumes which it stores in a CAD-format geometry file. In the second part, GPT and Geant4 are launched in linked Unix windows to simulate the GPT input file. The third part imports the simulation results into analysis and visualization programs.

Figure 1 shows the flow diagram for the first part where the python program *gptgeom2gdml* reads the GPT input file, *run1.in*, containing the GPT scattering elements and translates them into volumes in the *GDML* geometry language [3] and writes the file, *run1.gdml*. The 3D interpretation of the surfaces defined in the GPT can be viewed by importing the *run1.gdml* into the open-source CAD program FreeCAD [4].

Converting the GPT surfaces to 3D volumes requires picking thicknesses for the zero-thickness surfaces defined by GPT. In this work the geometry conversion program, *gptgeom2gdml.py,* uses 4 mm. Other thicknesses can be specified by editing the *.gdml* file, taking care not to violate GPT's surface definitions. Generally, it's prudent to confirm the geometry is correct by viewing it in FreeCAD.

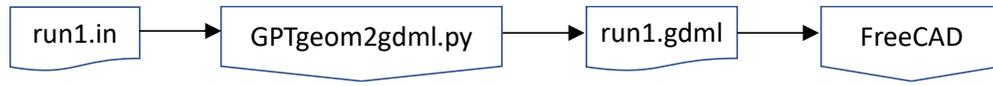

Figure 1. Conversion of the surfaces defined in GPT into solids using *gptgeom2gdml.py*. The solids CAD file has the extension *.gdml* and can be imported into FreeCAD or equivalent CAD program.

The next two parts of the system: simulation and data analysis, are flow charted in Figure 2. The GPT and Geant4 processes run asynchronously in linked Linux console windows, sharing memory and communicating with semaphores. The simulation begins with launching *GPTScattererGDML* in one Linux window after clearing the shared memory. *GPTScattererGDML* reads in the *.gdml* file and then pauses, waiting for GPT to semaphore that an electron has crossed a material boundary. Next GPT is launched in a separate window and the tracking simulation begins. If an electron crosses a material boundary, GPT semaphores the Geant4 process and waits for a reply. Then *GPTScattererGDML* reads the crossing electron's parameters from the shared memory and begins propagating the electron and depositing its energy into the material. Geant4 finishes and semaphores GPT when the electron has either stopped in the material (cutoff energy 400 eV) or re-crosses a boundary back into the vacuum. If the electron has stopped, GPT logs its position and deposited energy as a stopped electron and goes onto the next electron. If a boundary is crossed back into the beampipe, the Geant4 signals GPT to use the shared memory parameters to re-launch and track the secondary electron. This back-and-forth process continues until all the electrons either stop (fall below a threshold energy of 400 eV) or the simulation ends.

The results of the simulation are saved in standard GPT data files with extensions *.gdf* and can be analyzed with standard GPT tools. Alternatively, the data can be converted to ASCII and exported to MATLAB and other utilities for further analysis.

Since there is no provision in GPT for defining the material type, this information is passed to Geant4 by adding *{* after the *#* comment symbol, e.g., stainless steel is specified with: *#{'material':'G4_STAINLESS-STEEL'}*. *gptgeom2gdml.py* looks for these lines in the GPT-input file and uses them to set the material type in the *.gdml* file. The default material type is stainless steel. This usage is illustrated in the GPT input file for calculating the albedo. (See Appendix).

Besides the usual GPT analysis tools, an ascii file of the GPT trajectory data can be read into the 3D CAD program, FreeCAD, using a python macro. The python macro, *insert_tracks.py*, enables FreeCAD to plot each electron's path along the beamline, to view if and where it is scattered by beamline components. In addition, this macro can tag a group of trajectories at a 'screen' located along the beamline and highlight their tracks back to their sources.

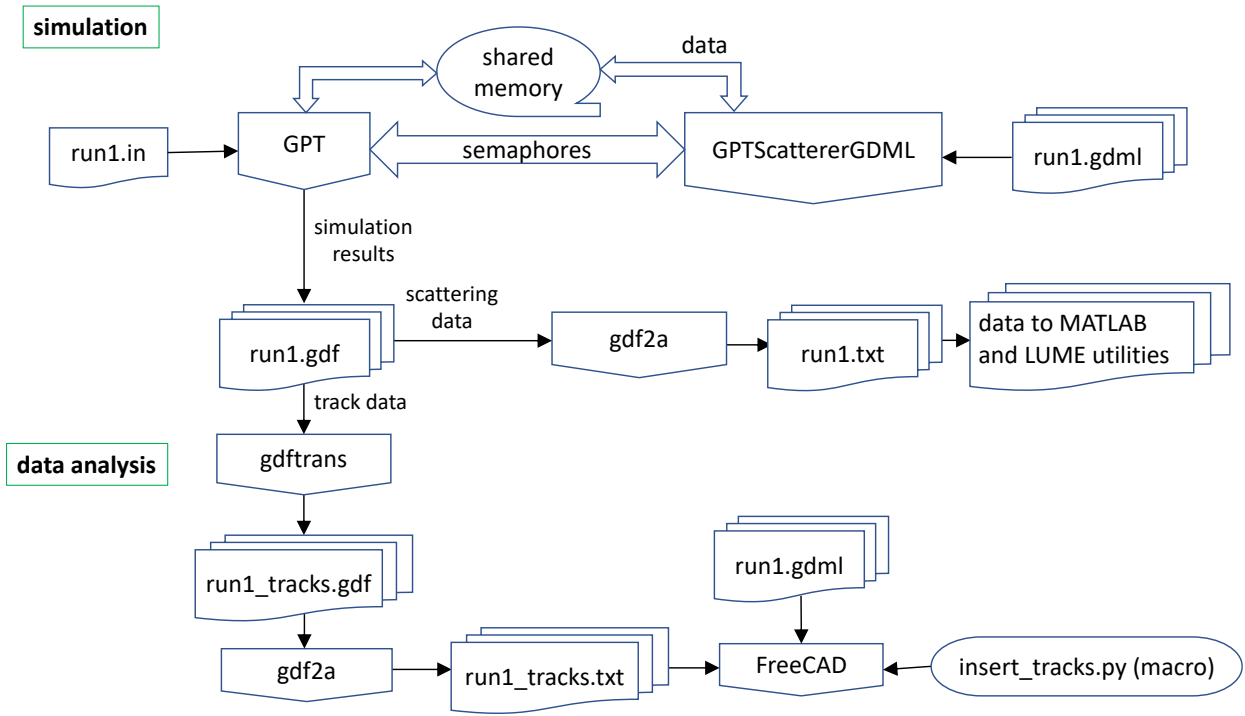

Figure 2. Flow diagram of the GPT+Geant4 programming. GPT and Geant4 communicate using semaphores and exchange data via shared memory. GPT, gdftrans and gdf2a are GPT programs, see GPT User Manual [1] for details.

### III. General Properties of Secondary Emission and GPT+Geant4 Energy Range

Geant 4 allows for a variety of options for controlling the electron-scattering process. In GPT+Geant4 we use the continuous-slowing-down approximation (CSDA) for the energy loss and small angle scattering, and Mott/Rutherford nuclear scattering for large angle scattering. Chapter 3 of the book by Reimer [5] contains an excellent description of the dynamics of low energy electron scattering. This section reviews the key properties of secondary emission, describes the processes modeled by GPT-Geant4 and discusses its energy range of applicability.

An energetic electron incident upon a surface produces secondary electrons whose generation depends upon the primary electron's energy and angle-of-incidence. The secondary emission yield, $\delta$, is defined as the number of secondary electrons emitted from a surface per incident primary electron. For metals the secondary yield peaks between 0.5 (beryllium) and 1.5 (gold) at primary energies of 200 eV and 800 eV, respectively. Figure 3 shows the general trend of $\delta$ as a function of the primary electron energy, $E_p$. The secondary yield peak is at primary electron energy $E_{pmax}$ and crosses unity yield at $E_I$ and at $E_{II}$. Ref [6] gives these three energies for a variety of materials. For copper, $E_I = 200$ eV and $E_{II} = 1500$ eV with similar values for iron, the principal constituent of stainless steel.

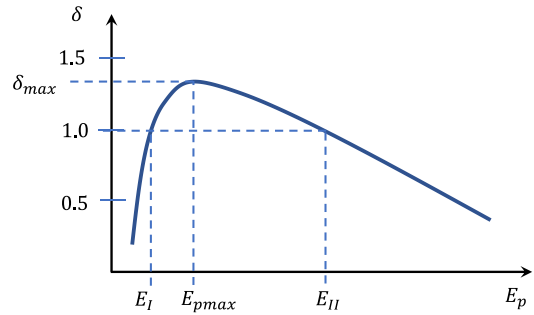

Figure 3. Secondary emission yield as a function of the primary electron energy. See Refs. [6,7].

The secondary yield is smaller than unity for primary energies much greater than $E_{II}$, and can be described by assuming just two energy loss processes: the continuous slowing down approximation (CSDA) and atomic nuclear scattering. CSDA accounts for small angle scattering and overall energy loss. Nuclear scattering, given by the Mott cross section, is responsible for large angle scattering.

These assumptions work well at 10 keV and higher, however they begin to fail as the primary energy nears $E_{II}$. For example, a GPT-Geant4 calculation of copper's secondary yield (aka the isotropic albedo) at $E_{II}$ should be 1 (by definition) but instead GPT-Geant4 gives 0.58. Despite this difference at 1 keV, the next section shows there is good agreement with measurements and published simulations for 10 keV (~10-times $E_{II}$) and higher primary energies. This agreement validates these assumptions in this energy region and the GPT-Geant4 simulations should accurately simulate the 800 keV portion of the LCLS-II injector. At higher energies multi-step processes like bremsstrahlung followed by pair production can generate electron and positron secondaries. The G4EmLivermorePhysics model used in the Geant4 simulations includes these and many more electromagnetic interactions [2]. But although many EM processes are computed by Geant4, this version of GPT+Geant4 tracks only the highest energy scattered electron.

### IV. Computing the albedo and validating GPT+Geant4

The scattering dynamics as they depend upon an incident electron's energy, angle-of-incidence and the material can be summarized in the backscattering coefficient, $\eta$, or the scattering albedo [8]. The albedo depends not only upon the material but also on the incident energy and the angle-of-incidence. The electron scattering albedo is defined as the ratio of the number of electrons scattered back out of the material to the number of incident electrons,

$$\eta = \frac{N_{scattered}}{N_{incident}}$$

As a test of its accuracy, GPT+Geant4 is used to calculate the albedo of a flat, semi-infinite plate and the results compared with published albedos. In the simulation, the plate is 1 km by 1 km and 4 mm thick which is thicker than the path length of the electrons considered here. The plate is made very large to intercept electrons at angles of incidence close to grazing incidence. The initial radial distribution of the incident electrons is uniform with a hard-edge radius of 0.1 microns. Typically, 10k electrons were tracked in each run with the albedo given by the number of backward-traveling electrons ($\beta_z < 0$) divided by the number of incident electrons. The initial beam is launched from z=0 with zero divergence and the plate is at z=10 cm.

Below are selected lines from the GPT input file given in the Appendix where *geant4* is defined as a custom scattering function. Comment lines in GPT begin with the # symbol. The first subroutine calls the new C-code, *geant4scatter*, defining the custom scattering function "*geant4*". The next line defines the scatterer to be stainless-steel with the *#{'material':'G4_STAINLESS-STEEL'}* command. As explained earlier, this command line is invisible to GPT but the combination of characters, #{, sets the material type for the Geant4 calculation, enabling a feature otherwise not available in GPT. After this is the *scatterplate* command defining the surface boundary as a 1kmx1km plate at z=10cm and specifies that the *geant4* scattering function be used. Note that since the *scatterplate* command does not give a thickness for the plate, 4 mm is assumed. (A different thickness could be specified by adding the key-value pair 'thickness': 'xxx' into the #{} dictionary; xxx is in mm.) The Appendix gives the rest of the GPT input file defining the beam, beamline components and specifying what data to save.

*# custom scattering function*
*geant4scatter("wcs", "I", "geant4");*
*#{'material':'G4_STAINLESS-STEEL'}*
*#scattering plate: 1km x 1km at z=10 cm*
*scatterplate("wcs","z",0.1,1000.0,1000.0)scatter="geant4";*

Figure 4 shows the primary (red) and secondary/scattered (blue) electron tracks for 800 keV electrons incident at 80 degrees onto a stainless-steel plate. These trajectories were calculated using the GPT input file given in the Appendix.

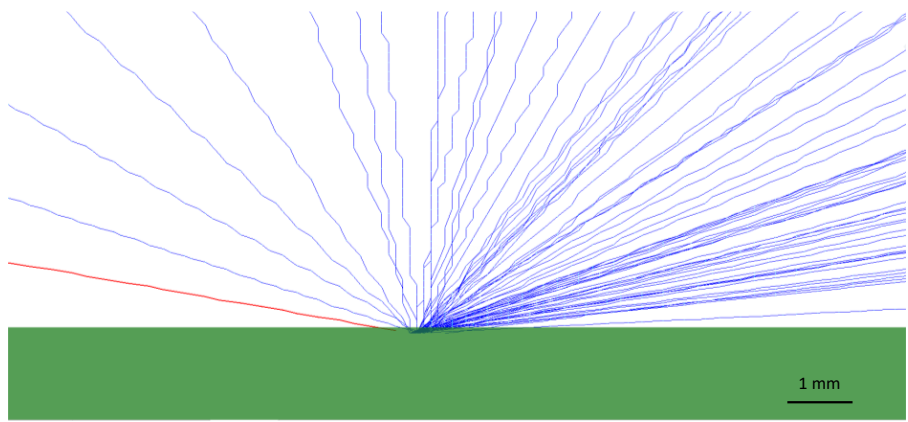

Figure 4: Side view of 800 keV electrons (red) at 80-degree incident angle scattering from a stainless-steel plate. The scattered electrons are drawn in blue. This simulation was for 100 incident electrons which produced 66 scattered electrons, for an albedo of 0.66.

### V. Validating the GPT+Geant4 code

Geant4 is a well-documented code and its results have been validated with measurements and comparisons with other simulation codes. The same is true for GPT which is well-known in the accelerator physics community. However, this is a new application combining both codes and what needs to be tested is the accuracy of the interaction between the two codes. Therefore, this section compares GPT+Geant4 calculated albedos with published measurements and simulations for a range of incident energies, angles-of-incidence, and material atomic numbers.

Figure 5 plots the albedo for aluminum as a function of the incident electron energy for normal incidence electrons. There is reasonable agreement between the albedo measurements and the GPT+Geant4 calculation over an energy range of two decades. The data shown are from Berger's seminal work in the 1960's [9].

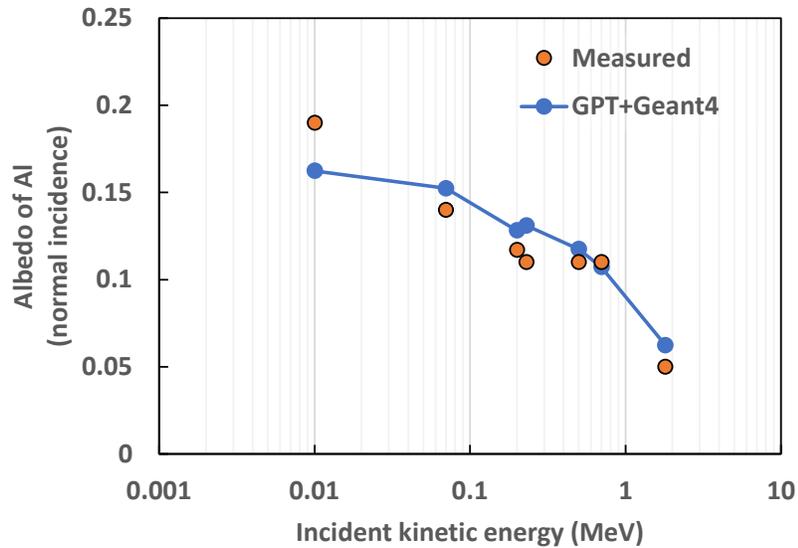

Figure 5. The albedo of aluminum vs. electron energy at normal incidence. The experimental measurements (orange points) are from the literature [9] and the GPT-Geant4 calculations are shown with blue line and points.

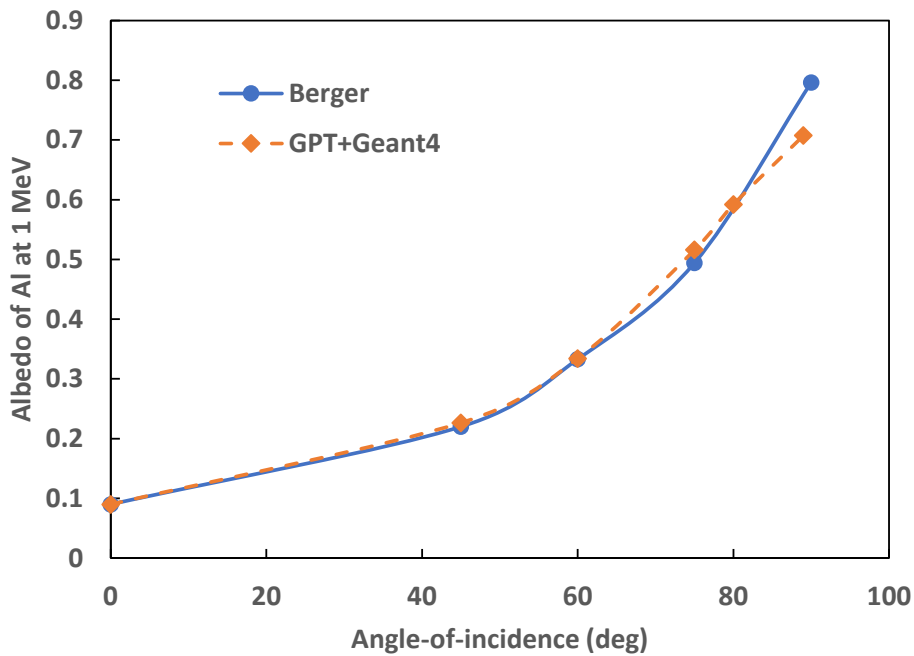

Figure 6. Albedo of aluminum vs angle-of-incidence for 1 MeV incident energy electrons. The blue curve with points is a simulation from Martin J. Berger, Ref. [9]. The orange points, dashed curve is the GPT+Geant4 calculated albedo as described in Section IV.

Figure 6 compares Berger's published simulation for the albedo of aluminum with the GPT+Geant4 calculated albedo. The agreement is quite good although nearly 60 years have passed between Berger's work and today's Geant4 calculations. However, the agreement is not surprising since Geant4 uses the same Monte Carlo methods as described by Berger in 1963. Both show the albedo increases a factor of 8 from normal incidence (0 degrees) to 85 degrees, near grazing incidence.

The measured and the GPT+Geant4 calculated albedos vs angle-of-incidence for 10 keV incident electrons are compared in Figure 8 for aluminum and copper. There is good agreement for copper, however the GPT+Geant4 albedo for aluminum is approximately 20% higher than experiment at 75 degrees while converging with the data at normal incidence. Given the demonstrated accuracy of Geant4 but risking hubris, we are tempted to believe the GPT+Geant4 albedos.

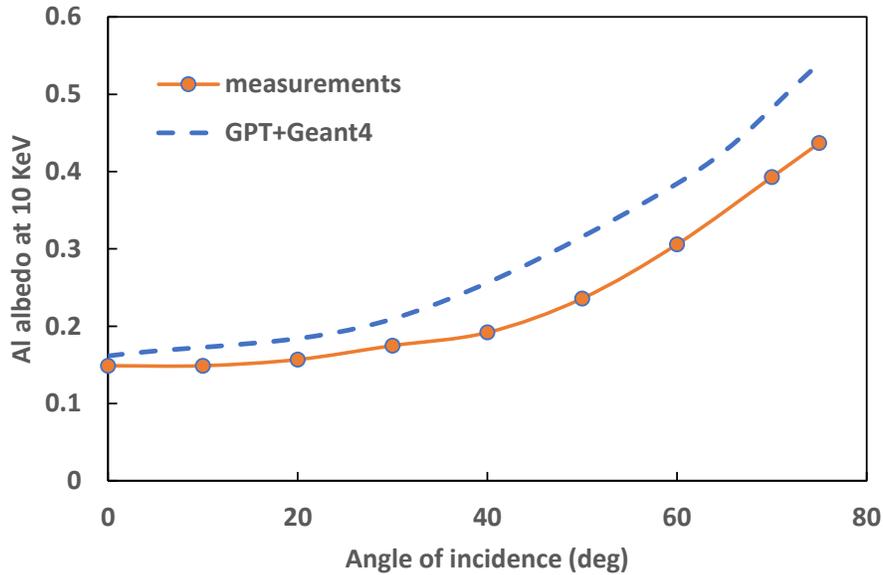

Figure 7. Comparison of the measured and calculated albedo of aluminum as a function of angle-of-incidence for 10 keV incident electrons. For measurements see Ref. [8].

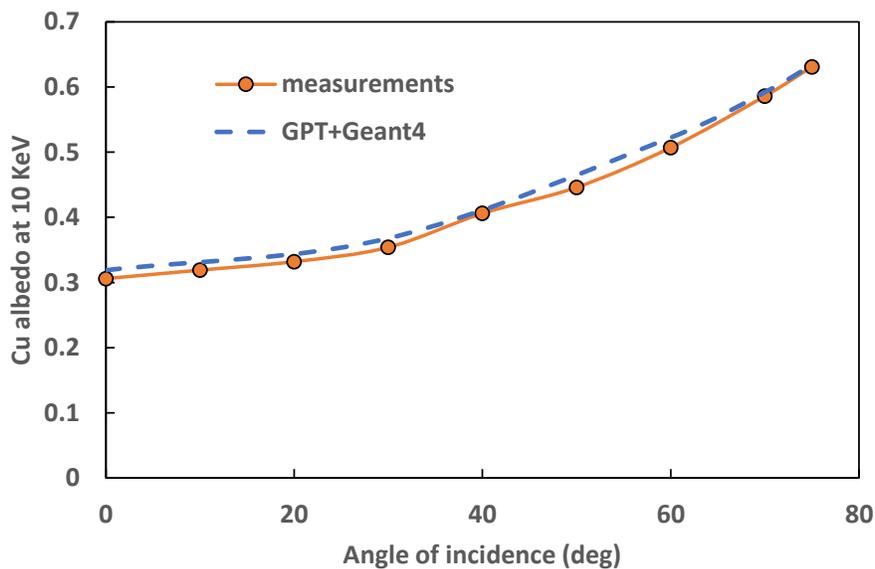

Figure 8. Comparison of the measured and calculated albedo of copper vs. angle-of-incidence for 10 keV incident electrons. For measurements see Ref. [8].

Because the beamline is fabricated from a variety of metals and plastics it is equally important to verify the GPT+Geant4 code is accurate for all elements across the periodic table. Figure 9 plots the normal-incidence albedo for the elements of the periodic table. Clearly GPT+Geant4 is in good agreement with the published data across the periodic table.

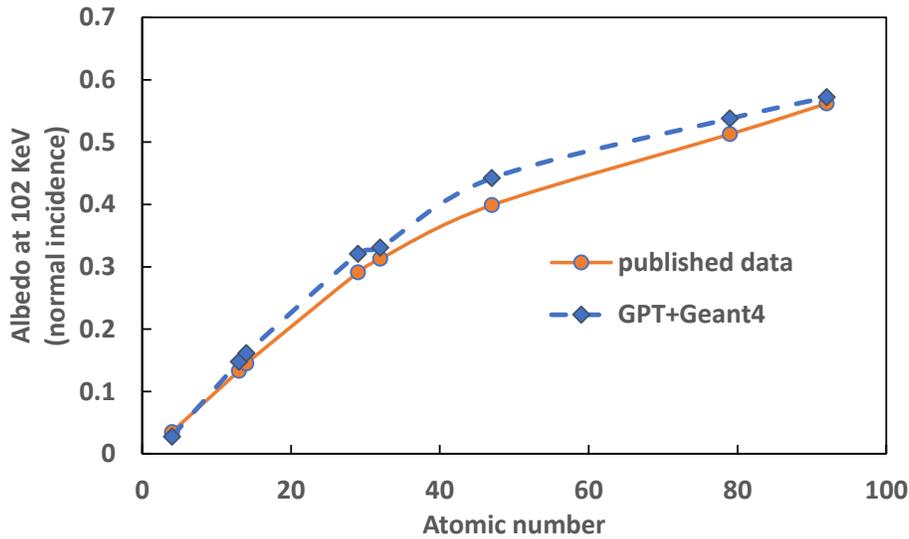

Figure 9. The normal incidence albedo vs the atomic number of the scatterer. For published data see Ref. [10].

And finally, in Figure 10, we compare the albedos of the three most used metals in an accelerator beamline: copper, stainless-steel and aluminum. For relevance, the albedo is plotted over the energy range of most electron guns from 10 keV to 2 MeV. The normal-incidence albedos of copper and stainless-steel are similar, while aluminum's albedo is less than half that of stainless-steel.

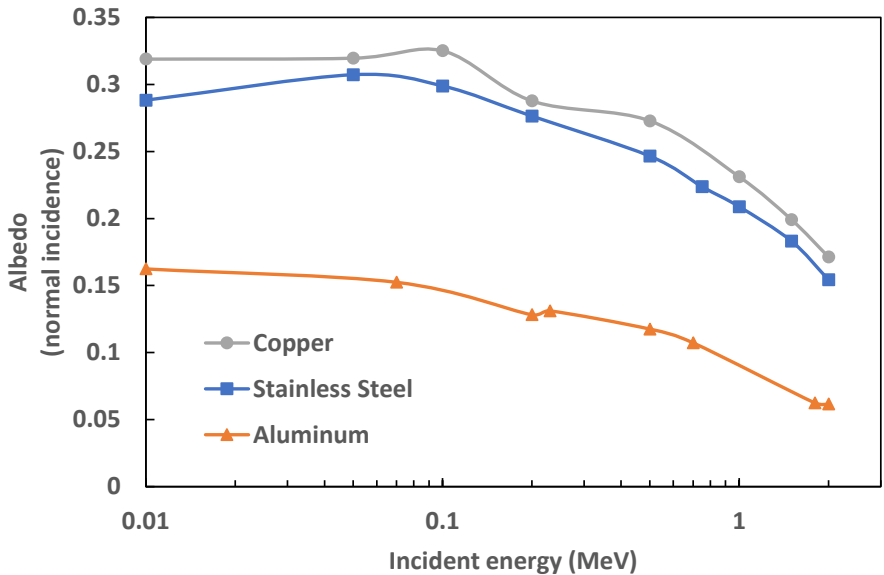

Figure 10. GPT+Geant4 values for the normal incidence albedo of copper, stainless-steel, and aluminum from 10 keV to 2 MeV. These are the three most used metals in electron injectors.

### VI. Analysis and visualization tools

As a real-world example illustrating the analysis and visualization tools developed for GPT+Geant4, we choose the LCLS-II injector which consists of a 187 MHz RF gun followed by two solenoids with a $7^{th}$ harmonic rf buncher between the solenoids [11]. The general layout is shown in Figure 9. The GPT input file and other files and information needed to run this example are given at the LUME-scatter github repository [12]. As outlined in Section II, *gptgeom2gdml.py* is used to generate a GDML file from the GPT input file. Then running GPT with our Geant4-developed application *GPTScattererGDML* produces the trajectories of the primaries and secondaries. FreeCAD draws the beamline geometry in 3D using the .gdml file and the *insert_tracks.py* macro then draws the trajectories.

The rf field shapes for the gun and buncher were calculated with Superfish [13]. A high-resolution grid of the gun's field was used to improve the simulation accuracy in the cathode-plug gap where the field emitters are located. The amplitude of the gun rf field was adjusted to give 700 keV gun kinetic energy as measured during commissioning. The buncher was on for these simulations with its field strength set to the injector design value [14]. The solenoid's on-axis Bz vs. z field shape was measured by SLAC magnetic measurements group [15]. The solenoid was set to the nominal integrated field used during commissioning, 0.0492 kG-m (6.05 A).

Figure 11 shows the 3D drawing of the first 1.6 meters LCLS-II beamline envelope along with the trajectories in different colors from four field emitters. The bottom left image shows the four emitters (trajectories shown with red, blue, green and magenta lines) around the circumference of the cathode plug and in the gap between the plug and the gun's nosecone. These are close to the four emitter locations observed during LCLS-II injector commissioning [16]. Each of the four emitters is assumed to emit 45 fC/RF cycle with a pulse shape and length as given in Ref. [17]. The emitters are located approximately 0.7 mm behind the cathode surface as can be seen in Figure 11 (lower left). The lower center image shows details of the electron trajectories near the collimator. The lower right-hand image gives one an end-on view of the trajectories looking back toward the cathode from the collimator.

These simulations confirm the 12 mm radius collimator is effective at blocking dark current generated at the edge of the cathode plug. The collimator is effective because the electron optics from the cathode to the collimator have a magnification of approximately three. This first-order magnification places the emitters' images at ~15 mm from the collimator center, making the 12mm-collimator effective and efficient. However, this also implies that any emission within a 4 mm radius of the cathode center will make it through the 12mm collimator. Similarly operating the solenoid at other fields will change the magnification and allow electrons through the collimator.

The GPT+Geant4 simulations give a collimation efficiency of 98% for the 12mm-collimator. Those electrons making it through the collimator have low energy and are strongly overfocused by the second solenoid (not shown) at z=1.65 meters and lost to the vacuum pipe walls.

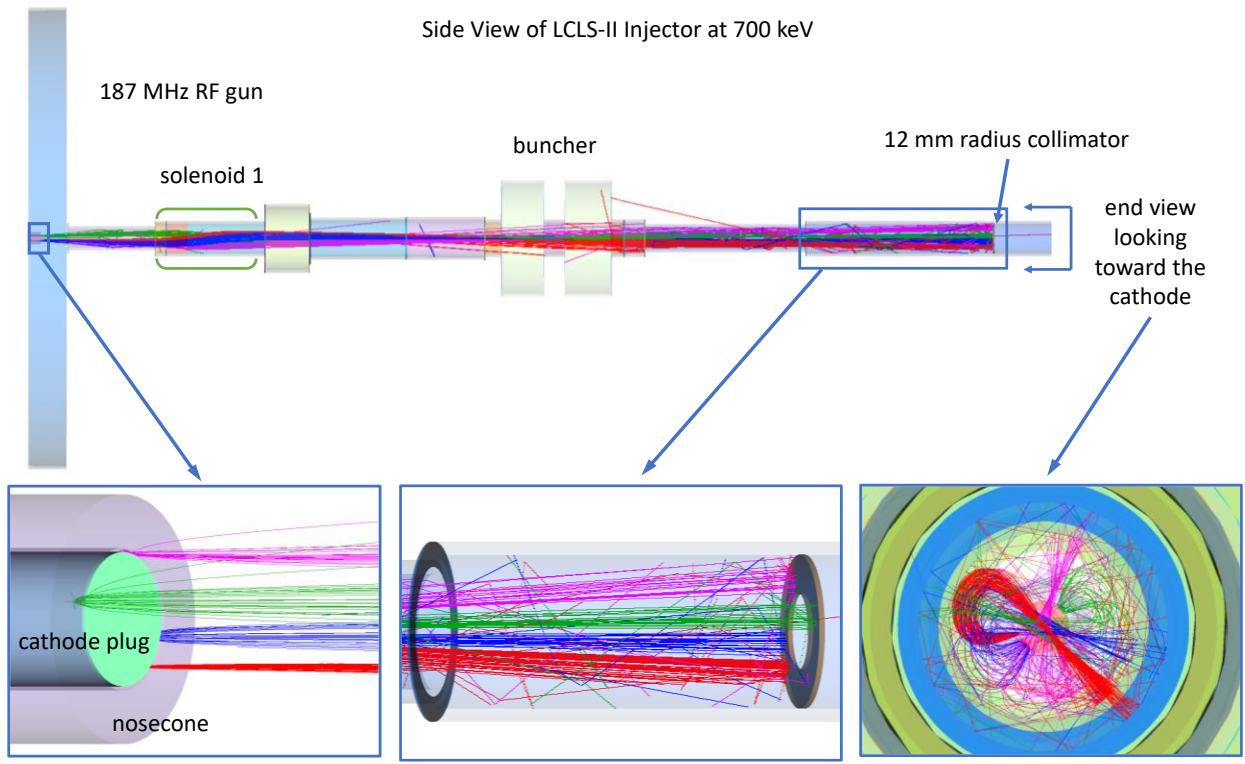

Figure 11. GPT+Geant4 dark current simulations of the LCLS-II injector. Top: 3D FreeCAD drawing of the beamline vacuum envelope with the primary and secondary trajectories from the photocathode to the 12 mm radius collimator at 1.5 m. Bottom (left): Detail of the emission from four emitters around the gap between the cathode plug and the gun's cathode nosecone. The cathode plug radius is 5 mm. Bottom (center): Detail of the primary and scattered electron trajectories on either side of the collimator. Bottom (right): View of the four emitters' trajectories looking from the collimator back toward the cathode.

The simulation data can also be analyzed using MATLAB and the python codes in GPT-LUME [18]. Figure 12 shows an example of a MATLAB-based analysis for the LCLS-II beamline. This is for the same case as shown in Figure 11 with four field emitters located around the circumference of the cathode plug.

Figure 12 shows scatter plots of the simulations for the four field emitters as described above. The four sources are approximately at the same orientations and positions as determined during LCLS-II injector commissioning [16]. The simulation assumes all four are located approximately 0.7 mm behind the cathode's surface. The red-colored electrons have been launched uniformly into all directions with 10 eV kinetic energy. The blue-, green- and magenta-colored electrons have been launched with 100 eV isotropic energy. The source radius is 1 micron for all four emitters.

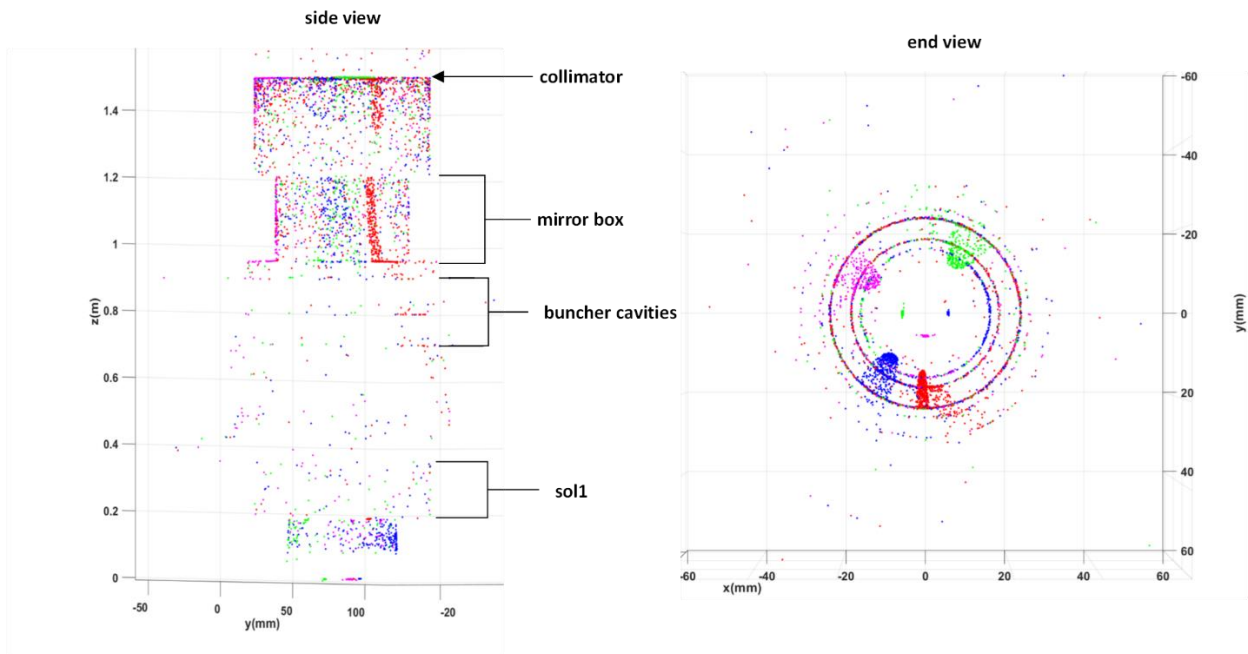

Figure 12. Scatter plots electrons hitting the beamline walls. Emission from four emitters in the gap between the cathode plug and nosecone at 5.022 mm radius is simulated. All emitters are 0.706331 mm behind the face of the cathode but have different isotropic energies. See text. Red points: 10 eV isotropic source energy. Blue, green, and magenta source points: 100eV isotropic source energy. Left: Side view of the electron hitting locations from the cathode to the collimator/YAG screen. Right: End view of the beamline hits.

The 3D-scatter plot on the left in Figure 12 is a side view of the LCLS-II injector beamline showing where electrons from these four emitters strike the walls of the beam pipe. The view extends from the cathode at z=0 (bottom) to the collimator at z=1.5 m (top) with the positions of the major beamline components.

The calculations assume the beampipe has rotational symmetry about the z-axis and uses the design dimensions for the beam pipe, the rf gun, the buncher cavities, the collimator, etc. [19]. Where the density is high, the striking electrons 'paint' the outline of the beamline walls. Because of its low source energy spread, the distribution of red-colored scattered points is transversely narrower than the other three (blue-, green-, and magenta-colors) which have 10-times the energy spread. However, because the longitudinal dynamics are dominated by the rf acceleration, the z-location of electrons striking the beamline is similar for the two source energies.

There is a small loss in the plug-nosecone gap which mostly hits the side of the cathode plug. The size of the loss increases with increasing launch energy. After that there are losses in the narrow pipe before the solenoid, but much reduced loss at the solenoid where the beam pipe has been enlarged, and beam loss remains low through the buncher cavities. However, the loss suddenly increases at the drive laser injection mirror box where the red-colored, 10 eV source energy electrons draw a narrow streak across the inside of the mirror box, and the other sources draw wider transverse swaths due to their larger energy spreads. There is then a short gap in the losses at z=1.2 meters where the pipe inner diameter is larger and shielded by the mirror box pipe. But the expanding beams again hit the walls near z=1.35 m and finally terminate at the collimator, z=1.5 m.

The scatter plot on the right in Figure 12 shows the end view of the electron losses looking down the injector's centerline toward the cathode from the collimator. The rotational symmetry can be seen in the ring structure. Near the center at 5 mm radius are the four sources, with the blue-emitter roughly at 3 o'clock, the red at 2:30, the green at 9, and the magenta at 6 o'clock. The density of the colors indicates how many electrons from each source strike the cathode plug. One sees fewer electrons from the low energy spread emitter (red) striking the plug. The heads of all four radial 'comets' are at 15mm radius when they reach the collimator while their low energy tails stretch out to larger radii becoming lost earlier along the beamline as seen in the left-hand scatter plot. The transverse width of each comet is proportional to the transverse energy spread at its source, and the radial length is related to the longitudinal energy spread. The large longitudinal energy results when the field emitter's long bunch length as it's accelerated in the gun's time-dependent rf field. And finally, the end view at the collimator shows the four beams all rotate the same ~115 degrees clockwise in the solenoid field.

One can easily see how this type of analysis and visualization could aid in designing beamline envelopes and interiors. In addition, these results suggest that measurements of the transverse beam size at the collimator/YAG could determine the isotropic energy of a field emission source. And in turn this energy would give the emitter's field enhancement factor and approximate geometry.

### VII. Summary and Conclusions

The goal of this work was to create a new tracking toolchain with realistic electron loss and scattering physics. The result is the merging of the well-established GPT and Geant4 codes, referred to as GPT+Geant4. In addition, interfacing with GPT+Geant4 is the novel capability to view in 3D the electron trajectories inside the beam pipe and to see where they come close to or hit the walls. This is done using open-source CAD software called FreeCAD. The addition of FreeCAD allows importing CAD files in a variety of formats and gives the user all the zooming, magnifying and panning features available with CAD software. In a sense, what we've developed could be called GPT+Geant4+FreeCAD simulation and visualization system of codes.

This paper presents the architecture and first results of our system of programs which unify the tracking of primary electrons and the secondaries they produce when they strike the beamline walls and other components. This was done by combining the well-known tracking code, GPT, with a well-documented and validated/trusted EM shower code, Geant4. Flow diagrams illustrate how these codes work together, using shared memory and semaphores, to track and scatter electrons. The industry standard geometry file format, *gdml*, was adopted to convert the surfaces defined by GPT into 3D volumes which can be used by Geant4 and imported into FreeCAD.

The GPT input file given in the Appendix was used to simulate the albedo of electrons scattering from a variety of materials, incident energies and angles-of-incidence. In all cases there is good to excellent agreement with the literature albedos, showing the combined system is working correctly.

The capabilities of the GPT+Geant4 analysis and visualization tools were demonstrated with the LCLS-II injector as an example. Field emission from four emitters located around the cathode plug's circumference was modeled from the cathode to a collimator 1.5 meters downstream from the cathode. Analysis of the GPT+Geant4 results show that 98% of the electrons coming from the cathode plug-nosecone gap are blocked by the 12-mm radius collimator at z=1.5 meters.

Further analysis was done in MATLAB using a custom program with imported simulation data. This analysis allowed easy viewing of the dependence of the transverse beam size upon the initial emitter energy. It was found that the emitters isotropic energy results in a transverse momentum spread which in turn affects the transverse distribution of the beam and its loss. That said, the z-dependence of the losses is similar for both large and small isotropic source energies.   The z-dependence is mostly a result of the beam's large energy spread and how it's focused and bent in the beamline optics.  Therefore, in both cases there is significant loss beginning with the mirror box and ending at the collimator.

The GPT-Geant4 programs with installation and operation instructions are freely available at the GitHub site given in Ref. [12].

## Acknowledgements

DHD wishes to thank Feng Zhou and Chris Adolphsen for their constructive criticism during the early stages of this work.  DHD was supported by the U.S. Department of Energy, Office of Science, under Contract No. DE-AC02-76SF00515. MH wrote the programs and donated his time.  SBG and MJL donated the GPT code and provided advice.

# Appendix
Listing of GPT input file for calculating the albedo.

```
#file:  albedo8.in
#albedo8: example in paper
#Initialize random number, particle mass, etc.
randomize(4763); #random number seed
mc2=0.511e6; #electron mass in eV
##Define beam properties
setparticles("beamA0",100,me,qe,-1e-12) ;  #setcharge to 1 pC for 100 particles
EA0=0.800e6; #starting kinetic energy in eV for beamA0
GA0=1+EA0/mc2;  # gamma of beamA0
GBA0=sqrt(GA0^2-1); #beamA0 beta*gamma
setGBzdist( "beamA0", "u", GBA0, 0 ) ; #start zero energy spread beam
setGBthetadist( "beamA0", "u", 0, 0.0 ) ;  #with zero divergence
setGBphidist( "beamA0", "u", 0, 2*pi/100 ) ;
##transverse dimensions of initial round, uniform beamA0
radius=0.0000001; # in meters
setrxydist( "beamA0", "u", radius/2, radius ) ;
setphidist( "beamA0", "u", 0, 2*pi ) ;
##
inangle=80*pi/180; #angle of incidence in radians
settransform("wcs", 0.000,  0.0, 0.00,    cos(inangle),0,sin(inangle), 0,1,0,"beamA0");#source at z=0cm
# Output control
tout(0); #turn off to get xscat, yscat, etc.
tstep=20e-12;
tout(0, 5e-9, tstep ) ; #turn off to get xscat, yscat, etc.
zminmax("wcs","I",-0.10,1.6);
#screens
screen("wcs", "I", 0.04);
screen("wcs", "I", 0.10);
screen("wcs", "I", 0.11);
# custom scattering function
geant4scatter("wcs", "I", "geant4");
#scattering plate: 1km x 1km at z=10 cm
#{'material':'G4_STAINLESS-STEEL'}
scatterplate("wcs","z",0.1,1000.0,1000.0)scatter="geant4";
```